\documentclass[usenatbib]{mn2e}
\usepackage{amsmath}
\usepackage{graphicx}
\usepackage{natbib}

\newcommand{\etal}{{et al.~}}
\newcommand{\lta}{\la}

\newcommand{\rmag}{\>^{0.1}{\rm M}_r-5\log h}

\newcommand{\kpc}{\>{\rm kpc}}

\newcommand{\beq}{\begin{equation}}
\newcommand{\eeq}{\end{equation}}

\newcommand{\msunh}{\>h^{-1}\rm M_\odot}

\newcommand{\apj}{ApJ}
\newcommand{\apjs}{ApJS}
\newcommand{\aj}{AJ}
\newcommand{\mnras}{MNRAS}
\newcommand{\aap}{A\&A}

\newdimen\hssize
\newdimen\hdsize
\begin{document}

\title[The Shape and Alignment of Dark Matter Haloes]
      {Probing the Intrinsic Shape and Alignment of Dark Matter Haloes
       using SDSS Galaxy Groups}

\author[Y. Wang et al.]
       {\parbox[t]{\textwidth}{
        Yougang Wang$^{1,8}$ \thanks{E-mail:wangyg@bao.ac.cn},
        Xiaohu Yang$^{2,3}$,
        H.J.~Mo$^4$,
        Cheng Li$^{2,3,5,6}$,
        Frank C. van den Bosch$^{7}$,
        Zuhui Fan$^8$,
        Xuelei Chen$^1$}\\
        \vspace*{3pt} \\
        $^1$National Astronomical Observatories, Chinese Academy
            of Sciences, Beijing 100012, China\\
        $^2$Shanghai Astronomical Observatory (SHAO), the Partner
            Group of MPA, Nandan Road 80, Shanghai 200030, China\\
        $^3$Joint Institute for Galaxy and Cosmology (JOINGC) of SHAO and USTC\\
        $^4$Department of Astronomy, University of Massachusetts, Amherst MA
            01003-9305\\
        $^5$Max-Planck-Institut f\"ur Astrophysik,
            Karl-Schwarzschild-Strasse 1, 85748 Garching, Germany\\
        $^6$Center for Astrophysics, University of Science and
            Technology of China (USTC), Hefei, Anhui 230026, China\\
        $^7$Max-Planck Institute for Astronomy, K\"onigstuhl 17,
            D-69117 Heidelberg, Germany\\
        $^8$Department of Astronomy, Peking University, Beijing 100871, China}

%%%%%%%%%%%%%%%%%%%%%%%%%%%%%%%%%%%%%%%%%%%%%%%%%%%%%%%%%%%%%%%%%%%%%%%%%%%%%%
\date{}

\maketitle

\label{firstpage}

%%%%%%%%%%%%%%%%%%%%%%%%%%%%%%%%%%%%%%%%%%%%%%%%%%%%%%%%%%%%%%%%%%%%%%%%%%%%%%

\begin{abstract}
  We study the three-dimensional and projected shapes of galaxy groups
  in  the Sloan Digital  Sky Survey  Data Release  4, and  examine the
  alignment  between the  orientation of  the central  galaxy  and the
  spatial   distribution  of   satellite   galaxies.   The   projected
  ellipticity of a group is measured using the moments of the discrete
  distribution of its member galaxies.  We infer the three-dimensional
  and projected axis  ratios of their dark matter  haloes by comparing
  the  measured  ellipticity distributions  with  those obtained  from
  Monte Carlo  simulations of  projected, triaxial dark  matter haloes
  with different axis ratios. We find that the halo shape has a strong
  dependence on  the halo mass.   While the haloes of  low-mass groups
  are nearly  spherical, those of massive  groups tend to  be prolate.
  For  groups  containing  at  least  four  members,  the  statistical
  distribution of their measured ellipticities does not have a strong
  dependence on  the colors of  their central galaxies.   Our analysis
  further  shows that  the  average three-dimensional  axis ratio  for
  haloes  with   $12<{\rm  log}[M/(h^{-1}M_{\odot})]\leq15$  is  about
  $1:0.46:0.46$, resulting  in a projected  axis ratio of $\sim  0.77$.
  Our results for the alignment between the orientation of the central
  galaxy of a  group and the distribution of  their satellite galaxies
  are  in broad  agreement with  those obtained  by Yang  et  al.  The
  distribution  of satellite galaxies  preferentially aligns  with the
  major axis  of the central galaxy,  with a clear  dependence on both
  halo  mass  and galaxy  colors.   In  particular,  the alignment  is
  stronger in more massive groups, and the strongest alignment is seen
  between red  centrals and the  distribution of red  satellites.  For
  groups  with blue centrals,  no significant  alignment is  detected.
  Finally,  we examine how  the observed  alignment can  be reproduced
  with  the  information about  the  halo  axis  ratios. The  observed
  alignment signal  can be reproduced  if the angle between  the major
  axis of the central galaxy and  the projected major axis of the host
  halo has  a Gaussian distribution with  a mean of  $0^{\circ}$ and a
  dispersion of $\sim23^{\circ}$. This dispersion is larger for groups
  with blue centrals than those with red centrals.
\end{abstract}


%111111111111111111111111111111111111111111111111111111111111111111
\section{Introduction}

In the cold dark matter  scenario, small dark matter haloes form
first and grow  subsequently to larger structures via  accretion and
merging processes.   Such processes  are  generally anisotropic  so
that  dark matter haloes  are expected to  be non-spherical. The
orientations of dark  matter haloes can  be related  to their
surrounding structures, such as  filaments and large-scale  walls
(e.g., Faltenbacher et al. 2002; Einasto et al. 2003; Avila-Reese et
al. 2005; Hopkins, Bahcall \& Bode 2005; Kasun \& Evard 2005;
Basilakos et al. 2006; Altay et al. 2006; Aragon-Calvo et al. 2006;
Maulbetsch et al. 2007; Ragone-Figueroa \& Plionis 2007; Hahn et al.
2007a, 2007b). The last major merger of a dark matter halo may play
an important role in determining its shape (van Haarlem \& van de
Weygaert 1993), although interactions between the gas and the dark
matter components are also excepted to play a role (Kazantzidis
\etal 2004).  Thus the shapes and orientations of dark matter haloes
contain abundant information about their formation histories, which,
in  turn depend on the underlying cosmology (Lee 2006; Ho et al.
2006).  On  the other hand,  the non-sphericity of dark matter
haloes can also lead to systematic errors in cosmological studies
(Sulkanen 1999; Wang \& Fan 2006). Therefore it is of great
importance to characterize the shapes of dark matter haloes,  both
observationally and theoretically.

The shapes  of clusters  of galaxies can  be probed using  X-ray
observations, studies  of the  Sunyaev-Zeldovich  effect, and
gravitational lensing  (e.g., Zaroubi et al. 1998, 2001; Relinsky
2000;  Lee \& Suto 2004; Wang \& Fan 2004; De Filippis et al. 2005;
Sereno et  al.  2006).  However, one can also use the spatial
distribution  of satellite  galaxies, since these  are expected  to
be good tracers  of the shapes of  their host haloes. The  large
redshift surveys carried out in recent years, for example, the
two-degree Field Galaxy redshift Survey (2dFGRS)  and Sloan  Digital
Sky Survey  (SDSS), have  provided angular positions and  redshifts
for  hundreds of thousands  of galaxies,  which allow detailed
studies  of the shapes of  large samples and the  shape dependence
on richness, multiplicity  and dynamical evolution of groups  and
clusters. Early studies preferred the prolate shapes (Carter \&
Metcalfe 1980; Binggeli 1980; Plionis, Barrow \& Frenk 1991; Fasano
et al. 1993; Cooray 2000; Basilakos, Plionis \& Maddox 2000; Orlov,
Petrova \& Martynova 2001), but did not  exclude the oblate
solutions. Plionis et al.(2004) (hereafter P04) estimated the shape
distribution  of UZC-SSRS2 groups  of galaxies by analyzing the
spatial distribution of group members. They found that the
prolate-like shape fits very well the cosmic structure on a large
scale apart from the disc galaxy. Somewhat surprisingly, they also
found that poor groups are more elongated than rich ones, results
which are opposite to what is found in numerical simulations (eg.,
Allgood et al. 2006; Kasun \& Evrard 2005).

 Somewhat surprisingly, these
results are opposite to what is found with numerical simulations
(Allgood et al. 2006; Kasun \& Evrard 2005). Recently, they
estimated the average group morphology of the Percolation-Inferred
Galaxy Group (2PIGG) and found that the prolate or, triaxial with
pronounced prolate shapes are the only acceptable morphological
model (Plionis et al. 2006).

In this paper, we use data from the SDSS to study the shapes of
galaxy groups, under  the assumption that  the spatial distribution
of group members  traces  the matter  distributions  of  their
underlying  dark matter haloes. Differing from previous  studies, we
run  Monte Carlo simulations to generate the same  number of member
galaxies as in each of the observed groups under  the assumption
that haloes of the groups follow the triaxial model of Jing \& Suto
(2002) (hereafter JS02). We then make  2D projected distributions of
these  member galaxies, which serve as  our Monte Carlo mock
samples.   We compare  the ellipticity distribution obtained  from
the mock  sample with the observed  one to extract information on
the axis ratios of dark matter haloes.

In the second part of this  paper we revisit the alignment between
the spatial distribution  of satellite  galaxies in  groups and  the
orientation  of their central galaxies (hereafter `centrals'). It is
important to assess accurately the alignment of the dark matter halo
because it holds important clues regarding the actual assembly
history of dark matter haloes. Extensive studies with high
resolution simulations have shown dark matter haloes have
anisotropic distributions of subhaloes that are aligned with their
major axis (Knebe et al. 2004; Libeskind et al. 2005; Wang et al.
2005; Zentner et al. 2005). This anisotropy mainly owes to a
preferred direction of satellite accretion along large-scale
filaments (Tormen 1997;  Vitvitska et al.  2002; Aubert, Pichon  \&
Colombi  2004; Knebe \etal 2004;  Wang et al. 2005; Zentner et al.
2005). In addition, the tidal forces from the host halo may also
induce new alignments (e.g., Ciotti \& Dutta 1994; Usami \& Fujimoto
1997; Fleck \& Kuhn 2003; Wang et al. 2007). Conversely, some
nonlinear effects such as violent relaxation and encounters can
weaken the primordial alignment (e.g., Porciani et al. 2002).

Comparing with the simulation studies, the observational search for
a possible alignment of central galaxies and their satellites has a
long and confusing history. The first study of such an alignment was
by Holmberg (1969), who found that satellites are preferentially
located along the minor axes of isolated disc galaxies. Holmberg's
study was restricted  to projected satellite-central distances of
$r_p  \lta 50 \kpc$. Subsequent studies, however, were unable to
confirm this so-called 'Holmberg effect' (Hawley \& Peebles 1975;
Sharp, Lin \& White 1979; MacGillivray  et al. 1982). Zaritsky et
al. (1997) studied the distribution of satellites around spiral
hosts and were also unable to detect any significant alignment for
$r_p \lta 200 \kpc$, but they found a preferred minor-axis alignment
for $300 \kpc \lta r_p \lta   500 \kpc$.  Our Milk Way and M31 have
satellites that lie in great planes that are highly inclined to
their discs (Lynden-Bell 1976, 1982; Majewski 1994; Hartwick 1996,
2000; Kroupa et al. 2005; Koch \& Grebel 2006; McConnachie \& Irwin
2006; Metz et al. 2007). With large redshift surveys, such as 2dFGRS
and SDSS, much larger samples  of galaxy groups can  be used to
discuss the alignment problem. Sales \& Lambas (2004) used a set of
1498 host galaxies with 3079 satellites from the 2dFGRS and found a
large-scale ( alignment of the satellites along the host minor axes
for $300 \kpc \lta r_p \lta 500 \kpc$. Brainerd (2005) studied  a
sample of isolated SDSS galaxies and found that the distribution of
satellite galaxies is strongly aligned with the major axis  of the
disc host galaxy. Yang et  al. (2006, hereafter Y06), using a galaxy
group catalogue similar to the one used here, but based on the SDSS
Data Release 2 (DR2), studied the alignment signal as function  of
the colors of the central and satellite galaxies.  They  found that
the alignment strength  is strongest between red  centrals  and red
satellites, while  the satellite distribution  in systems  with a
blue  central galaxy  is consistent with being isotropic.  Y06 also
found that the alignment strength is stronger in  more  massive
haloes  and  at smaller projected  radii from the  central galaxy.
These results have subsequently been confirmed by several
independent studies (Donoso, O'Mill  \& Lambas 2006; Azzaro  et al.
2007;  Agustsson \& Brainerd 2006a, 2007). Using the  same group
catalogue as that used here, Faltenbacher et al.  (2007a)  examined
several additional alignment signals. They found that the
orientations  of red satellites are preferentially aligned radially
in the direction of the brightest group galaxies (BGG). In addition,
they  found  a weak but significant indication that the orientations
of satellite galaxies are directly aligned with that of their BGG.
Comparing with the earlier studies of the alignment between
brightest cluster galaxies (BCGs) and their parent clusters (Carter
\& Metcalfe 1980; Binggeli 1982; Struble 1990; West 1994; Kim et al.
2001), Faltenbacher et al. (2007a) have given more detailed results
for the large samples. These various detections of alignment between
centrals and satellites have triggered a number  of investigations
into  the connection between the shapes and orientations of dark
matter haloes and their galaxy population, with the goal to improve
our understanding  of the formation  of dark  matter haloes and
galaxies (e.g.  Agustsson \& Brainerd 2006b, hereafter, AB06; Kang
et al. 2007, hereafter K07; Faltenbacher et al. 2007b; Brunino et
al. 2007; Sales et al. 2007; Pereira et al. 2007).

In addition to measuring the  alignment signals from the SDSS
observations for groups  of different  masses,  we try  to  infer
the  correlation between  the orientations of  the central galaxy
and  that of its  host halo, statistically but  both from the
observations. If  the projected  orientation of  a central galaxy is
perfectly aligned with the  projected orientation of  its host dark
matter halo,  assuming that satellite galaxies trace  the matter
distribution, the alignment  signal between the  distribution of
satellite galaxies  and the orientation  of  their central  galaxy
is strongest. By  comparing  the alignment signals measured from the
observations and the Monte Carlo samples, we  can estimate the
deviation  (misalignment  angle) of  the orientation  of central
galaxy  from the orientation  of its host  dark matter halo. We  use
a Gaussian distribution function to quantify this misalignment
angle.

This paper is organized as follows.  In \S2, we briefly describe the
observational data used for this study.  Section 3 presents our
measurements of the intrinsic shapes of dark matter haloes, where
the three-dimensional and two-dimensional axis ratios are determined
using Monte Carlo simulations. Section 4 shows the alignment signal
we measured from the SDSS and its implication for the shape
correlation between the central galaxies and the dark matter haloes.
Finally, \S5 presents a summary and discussion.
  Throughout this paper we refer to the inferred shape from the satellite
  galaxy distribution as the shape of the group and the corresponding
  dark matter halo, and use the major-axis direction of the
  satellite distribution to indicate the orientation of the group
  and the corresponding dark matter halo.

\section{Data}
\label{sec_data}

The analysis presented in this paper is based on the SDSS DR4 galaxy
group catalogue of Yang \etal (2007).  This group catalogue is
constructed applying the halo-based group finder of Yang \etal
(2005a) to the New York University Value-Added Galaxy Catalogue
(NYU-VAGC; see Blanton \etal 2005), which is based on SDSS DR4
(Adelman-McCarthy \etal 2006).  From this catalogue Yang \etal
selected all galaxies in the Main Galaxy Sample with redshifts in
the range $0.01 \leq z \leq 0.20$ and with a redshift completeness
${\cal C} > 0.7$. This sample of galaxies is used to construct three
group samples: sample I, which only uses the $362356$ galaxies with
measured redshifts from the SDSS, sample II which also includes
$7091$ galaxies with SDSS photometry but with redshifts taken from
alternative surveys, and sample III which includes an additional
$38672$ galaxies that lack a redshift due to fiber-collisions, but
which we assign the redshift of its nearest neighbor (cf.  Zehavi
\etal 2002). The present analysis is based on sample II which
consists of $369447$ galaxies distributed over 301237 groups with a
sky coverage of $~4514\, {\rm deg^2}$. Details of the group finder
and the general properties of the groups can be found in Yang et al.
(2007).

In this paper, the central galaxy is defined to be the brightest
galaxy in the group and other galaxies are satellites. We also take
the most massive (in terms of stellar mass) group member as the
central galaxy. As we have tested, the difference between these two
definitions is too small to be noticed. The group masses are
estimated using the ranking of group's characteristic luminosity,
$L_{19.5}$, defined as the combined luminosity of all group members
with ${^{0.1}}M_r-5\log h\leq-19.5$.  More details of the mass
estimations can be found in Yang et al.  (2007). Note that, the
selected galaxy groups contain a small fraction
  of interlopers, i.e. false members assigned to a group.
  If the distribution of these interlopers is uncorrelated (or
  anti-correlated) with that of the true members of the group,
  our results on both the ellipticity and the alignment can
  be biased. According to Yang et al.  (2007; 2005a) the
  average fraction of interlopers in the group is less than 20\%.
  We have tested the effect of such fraction by assuming that
  the distribution of the interlopers is uncorrelated with the shape of
  of the group and is spherical, we find that the presence of
  the interlopers can decrease the ellipticity of the groups and
  the alignment signals by $\sim 10\%$.

Note  that in  these group  catalogues survey  edge effects  have
been taken into  account (Yang  et al.  2007).   Only groups  with
$f_{\rm
  edge}\geq0.6$ are  selected, where $1-f_{\rm edge}$  is the fraction
of galaxies  in a group that are  missed due to the  edge effects.  In
order to obtain the ellipticity distribution of galaxy groups, we only
use groups with at least four members (one central galaxy and at least
three  satellites), which  results  in  a catalogue  of  5184 groups.
However, in studying the  alignment between satellite galaxies and the
orientation  of their  centrals, we  enlarge our  sample by  using all
groups  with at least  two members  (one central  and one  satellite).
This  sample gives a  total of  62212 unique  central-satellite pairs,
many more than in the DR2 sample used by Y06.

\section{The intrinsic shape of the dark matter halo}
\label{sec_shape}

\subsection{Methodology}

We now describe how we use the satellite distribution to determine the
ellipticity distribution of galaxy groups and their corresponding dark
matter haloes. The observed  satellite distribution in a group suffers
from  severe discreteness  effects.  In  particular, since  each group
contains only  a small number  of galaxies, there  is a high  level of
Poisson noise  in the  determination of the  ellipticity based  on its
galaxy distribution.  Thus the  ellipticity directly measured can only
be used as a rough  indicator of the underlying, true ellipticity.  We
will  use  mock  samples  to  quantify how  the  observed  ellipticity
distribution is  related to the real distribution.   Assuming that the
distribution  of  satellite  galaxies  in  a  group  traces  the  mass
distribution in the corresponding dark matter halo, we can infer, in a
statistical sense, the shapes of  dark matter haloes from the observed
distribution  of the  group  ellipticities.  In  order  to obtain  the
principal axes and the orientation of a group projected on the sky, we
define the inertia tensor as
\begin{equation}
X_{ij}=\sum_{i=1}^nx_{i,n}x_{j,n}
\end{equation}
where  $(x_{i,n}, x_{j,n})$  are the  projected coordinates  (with
the central galaxy at  the origin) of the $n^{th}$  satellite
galaxy. The semi-major and semi-minor axes of the ellipse, $L_a$ and
$L_b$ (two roots of the following equation), can be derived by
solving the equation
\begin{equation}
\left| \begin{array}{cc}
X_{11}-L^2 & X_{12} \\
X_{12}     & X_{22}-L^2 \\
\end{array} \right|
= 0 \; \; \mbox{.}
\end{equation}
The   direction   of   the   major   axis  is   given   by   the   eigenvector
${\textbf{r}}=[1,(L_a^2-X_{11})/X_{12}]$,  while the  ellipticity, $\epsilon$,
and the axis ratio, $\eta$, are
\begin{equation}
\epsilon=1-L_b/L_a,~~~\mbox{and}~~~~ \eta=L_b/L_a\,.
\end{equation}
Throughout this paper, we use the ellipticity $\epsilon$ and axis ratio $\eta$
to refer to the quantities {\it measured directly} from the data. The inferred
shapes of the  underlying dark matter haloes are described  either by their 2D
or 3D axis ratios.

As mentioned above, to quantify the true shapes of the dark matter haloes
associated with the galaxy groups, one needs to construct mock samples that
include the same selection effects. We construct Monte Carlo SDSS DR4 group
catalogues as follows.  First, we determine the number of satellites for each
of the SDSS groups.  Second, we re-distribute these satellites according to a
spherical NFW profile (Navarro et al. 1996, 1997) or a triaxial profile by
JS02.  Third, we project the three dimensional distribution of satellite
galaxies onto a two dimensional plane and measure the ellipticity,
$\epsilon_{MC}$, for each Monte-Carlo group. Such Monte Carlo approach was
firstly introduced by Basilakos, Plionis \& Maddox (2000) to recover the
projected cluster ellipticity distribution and the true projected ellipticity
distribution taking into account the background and discreetness effects (see
also Plionis et al. 2006). Thirty Monte-Carlo realizations are generated, and
we estimate the average and scatter of the ellipticity distribution using
these thirty Monte Carlo samples.  Some small groups with masses smaller than
$10^{11.6}\msunh$, where SDSS DR4 group catalogue does not provide mass
estimates, are removed from our sample.

The  first  model  for  the  mass  profile used  in  our  Monte
Carlo simulations is a spherical NFW density profile,
\begin{equation}
\rho_{\rm NFW}(r)=\frac{\rho_0\delta_c}{(r/r_s)(1+r/r_s)^2}
\end{equation}
where $\rho_0$ is the average mass density of the universe, $r_s$ is a
scale  radius  and  $\delta_c=200c^3/3[{\rm  ln}(1+c)-c/(1+c)]$.   The
concentration  parameter  $c$  is  defined  as  $c=r_{200}/r_s$,  with
$r_{200}$ the  radius within which the  mean density is  200 times the
average mass density of the  universe.  This model is fully determined
for  a halo of  a given  mass (or,  equivalently, $r_{200}$)  once the
concentration parameter is  given. The concentration parameter depends
on the halo mass  $M$ and redshift $z$, for which we  use the model of
Bullock et al.  (2001):
\begin{equation}
c(M,z)=\frac{c_\star}{1+z}\bigg(\frac{M}{10^{14}h^{-1}M_{\odot}}
\bigg)^{-0.13}\,,
\end{equation}
where $c_{\star}=8$,  as is appropriate for the ${\rm \Lambda CDM}$
model.

The  other density profile  we use  is that  proposed by  JS02.  Using
high-resolution  numerical  simulations,  JS02  proposed  an  NFW-like
triaxial density profile for dark matter haloes, which has the form
\begin{equation}
\rho(R)_{\rm JS02}=
\frac{\rho_0\delta_c}{(R/R_0)^\alpha(1+R/R_0)^{3-\alpha}}\,,
\end{equation}
where  $R=a(x^2/a^2+y^2/b^2+z^2/c^2)^{1/2}$,  and  $a\geq  b\geq  c$
are  the lengths of  the three principal semi-axes.   For the value
of  $\alpha$, it is found that  both $\alpha=1$ and  1.5 can provide
a good fit to  the simulated profiles. Detailed  comparisons showed
that $\alpha=1$ is  slightly better for cluster-scale haloes, while
$\alpha=1.5$  gives better fit for galactic haloes (JS02).
Therefore, we  adopt $\alpha=1.5$ for groups with  masses smaller
than $10^{13}\msunh$,  and  $\alpha=1.0$ for  more  massive  groups
(In fact,  the results  are not  sensitive to  the value  of
$\alpha$  adopted. We  have also adopted $\alpha=1.0$ for all
groups, their differences are tiny.).

\subsection{Ellipticity distribution}

\begin{figure}
  \resizebox{\hsize}{!}{\includegraphics{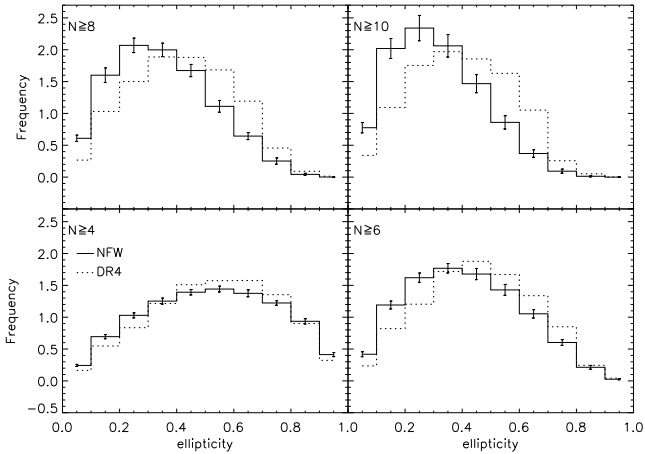}}
  \caption{Probability  distributions of the  ellipticity in  the SDSS
    DR4  (dotted  histogram)   and  Monte  Carlo  re-samplings  (solid
    histogram with error bars).   Here we assume that the distribution
    of satellite galaxies in groups follows a spherical NFW  profile.}
  \label{fig:NFW}
\end{figure}

\begin{figure}
  \resizebox{\hsize}{!}{\includegraphics{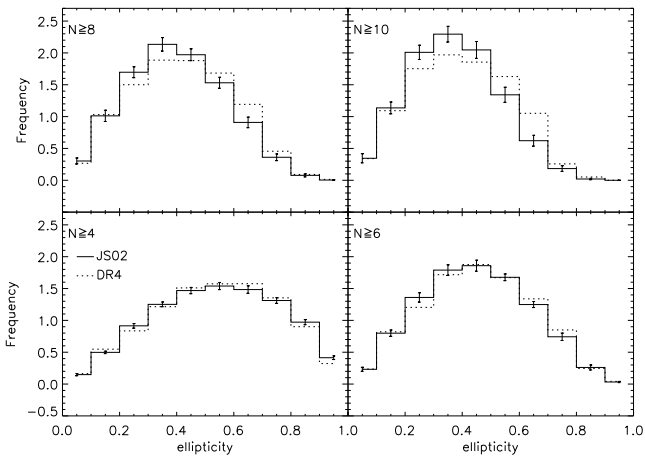}}   \caption{Same  as
    Fig.~\ref{fig:NFW}, but here we assume that the satellite galaxies
    are  distributed in  groups  according to  JS02  triaxial model.}
  \label{fig:Tri}
\end{figure}

Figs.~\ref{fig:NFW}  and \ref{fig:Tri} show  the probability  distributions of
the ellipticity  obtained from  the SDSS DR4  (dotted histogram) and  from the
Monte   Carlo   simulations   (solid    histogram   with   error   bars).   In
Fig.~\ref{fig:NFW},  we assume  that  the distribution  of satellite  galaxies
follows a  spherical NFW profile,  while in Fig.~\ref{fig:Tri} we  assume that
the distribution  of satellite galaxies follows  the JS02 model.  For the JS02
triaxial distribution, we  have adopted the model parameters  (axis ratios and
concentrations)  given   in  JS02.  In   both  cases  we  randomly   select  a
line-of-sight direction, project  the three-dimensional satellite distribution
onto a  two-dimensional plane, and measure the  ellipticity distribution using
the method  outlined above.   In each plot,  the parameter $N$  represents the
lower limit  on the number  of galaxies in  each group (including  the central
galaxy).  Note that the ellipticity distribution is strongly dependent on this
lower limit,  with poorer  groups being more  elongated. The  mean ellipticity
increases from $\sim0.40$ for groups  with richness $N\ge10$ to $\sim0.54$ for
groups with $N\ge4$.  However, this does not mean that the  true halo shape is
more  elongated for  poorer  groups; the  trend  is largely  a  result of  the
discrete sampling.  For example, in  the extreme case where only one satellite
galaxy is observed,  the measured ellipticity will always  be unity regardless
of  the shape of  the underlying  dark matter  halo.  Thus,  the shape  of the
underlying  halo  shape can  only  be  probed in  an  indirect  way, i.e.,  by
comparing the observed distribution with that of the Monte-Carlo samples.

From Fig.~\ref{fig:NFW} and \ref{fig:Tri}, it appears that the
triaxial model of JS02 fits the data better relative to the
spherical NFW model.

\subsection{The three-dimensional and projected shapes of dark matter haloes}

Numerous  studies  have  used  numerical  $N$-body simulations  to
probe  the non-spherical shapes of dark matter  haloes as traced by
dark matter particles (e.g., JS02; Kazantzidis et al.   2004), or by
subhaloes (e.g., Diemand, Moore \&  Stadel 2004).  The subhaloes are
more  closely associated  with satellite galaxies,  which   have
been  found  to   be  biased  tracers   of  the  mass distribution.
There  is  a  negative  spatial  bias  at  the  center,  and  a
corresponding  positive  velocity bias.   It  is  still  unclear
whether  this reflects numerical artefact (i.e., overmerging), or
whether this is real. Yang et al.  (2005b) studied the satellite
distributions in the  2dFGRS groups and found  evidence that  the
number  density distribution  of satellites  is less concentrated
than expected dark matter. Here  we probe the mean values of axis
ratios of dark matter haloes as traced by SDSS galaxies.  For this
purpose, we first  divide the  SDSS DR4  groups into  subsamples
according  to  their halo masses, and  measure the corresponding
ellipticity distribution as  traced by the satellite galaxies. Then,
using the model of JS02 with given axis ratios, we  construct thirty
realizations of  Monte Carlo  simulations and  measure the
corresponding ellipticity  distributions.  By changing  model
parameters (i.e. axis  ratios)  so  that  the  predicted ellipticity
distributions  match  the observed one, we  determine the underlying
axis ratios  of dark matter haloes. Here the comparison between
model predictions and observation is done in terms of a $\chi^2$,
defined as
\begin{equation}\label{eq:chi2}
\chi^2=\sum_{i=1}^{N_b}\frac {(\langle f_{i(MC)}\rangle
-f_{i(obs)})^2}{\sigma^2(f_{i(MC)})}
\end{equation}
where    $N_b=10$    denotes   the    bin    number    of   the
ellipticity distribution.   $\langle  f_{i(MC)}\rangle$  and
$\sigma^2(f_{i(MC)})$  are, respectively,  the   average  amplitude
and  1-$\sigma$  deviation   of  the ellipticity distributions
obtained from the thirty realizations  of Monte Carlo simulations,
while $f_{i(obs)}$  is   the  amplitude  of   the  ellipticity
distribution obtained  from the SDSS  groups. Note that  for each
set  of axis ratios, $\sigma^2(f_{i(MC)})$  changes slightly.
However, as we  have tested, using constant $\sigma^2(f_{i(MC)})$
does not have a significant impact on our measurement  of the
best-fit  axis ratios.  Thus we  use $\sigma^2(f_{i(MC)})$ computed
from the thirty Monte Carlo simulations and estimate the best-fit
axial ratios by minimizing  the $\chi^2$. Here the minimization  is
performed on the regular $100\times 100$ (3D) or $100$ (2D) axis
ratio grids.

\begin{table}
 \caption{Best-fit parameters for groups in different halo mass bins.
 From top to bottom, values are listed for groups with different members
 $N\geq4$, 6, 8, and 10, respectively.}\label{table:ratios}
%\begin{center}
\begin{tabular}{lccccc}\hline
  Mass bin & 3D axis ratios & $\chi^2$ & 2D axis ratios & $\chi^2$ \\
           & (I)            & (I)      & (II)           & (II)
  \\ \hline\hline
  $12<M^{\prime}\leq13^a$&1:0.96:0.85&22.58&$0.96^{+0.01}_{-0.02}$&46.16\\
  $13<M^{\prime}\leq14$  &1:0.80:0.72&29.93&$0.83^{+0.02}_{-0.01}$&28.30\\
  $14<M^{\prime}\leq15$  &1:0.88:0.44&11.86&$0.69^{+0.01}_{-0.01}$&19.13\\
  $12<M^{\prime}\leq15$  &1:0.46:0.46&67.63&$0.77^{+0.03}_{-0.01}$&56.02\\
 \hline
  $12<M^{\prime}\leq13$  &1:0.96:0.95&12.47&$0.86^{+0.02}_{-0.04}$&20.49\\
  $13<M^{\prime}\leq14$  &1:0.96:0.63&10.75&$0.76^{+0.01}_{-0.01}$&16.29\\
  $14<M^{\prime}\leq15$  &1:0.35:0.35&17.78&$0.68^{+0.01}_{-0.01}$&52.10\\
  $12<M^{\prime}\leq15$  &1:0.51:0.48&13.17&$0.72^{+0.01}_{-0.01}$&14.63\\
 \hline
  $12<M^{\prime}\leq13^b$&--&--&--&--\\
  $13<M^{\prime}\leq14$  &1:0.87:0.53&5.36&$0.72^{+0.01}_{-0.01}$&10.40\\
  $14<M^{\prime}\leq15$  &1:0.33:0.33&33.19&$0.70^{+0.01}_{-0.02}$&89.14\\
  $12<M^{\prime}\leq15$  &1:0.43:0.43&17.05&$0.72^{+0.01}_{-0.01}$&29.53\\
 \hline
  $12<M^{\prime}\leq13$  &--&--&--&--\\
  $13<M^{\prime}\leq14$  &1:0.84:0.50&7.68&$0.72^{+0.01}_{-0.01}$&12.63\\
  $14<M^{\prime}\leq15$  &1:0.61:0.36&34.84&$0.68^{+0.01}_{-0.01}$&60.24\\
  $12<M^{\prime}\leq15$  &1:0.39:0.39&23.04&$0.71^{+0.01}_{-0.01}$&48.66\\
 \hline

\end{tabular}
%\end{center}
 {\footnotesize
 \noindent
 $^{a}$ $M^{\prime}={\rm log}[M/(h^{-1}M_{\odot})]$. \\
 $^{b}$ The axis ratios in the $12<M^{\prime}\leq13$ mass bin with
        $N\ge8$ and $N\ge10$ are absent because our group sample
        contains too few of these rich groups for a reliable
        measurement of the ellipticity distribution.}
\end{table}

\begin{figure*}
%\centerline{\psfig{figure=fig3.ps,width=\hssize}}
  \resizebox{\hsize}{!}{\includegraphics{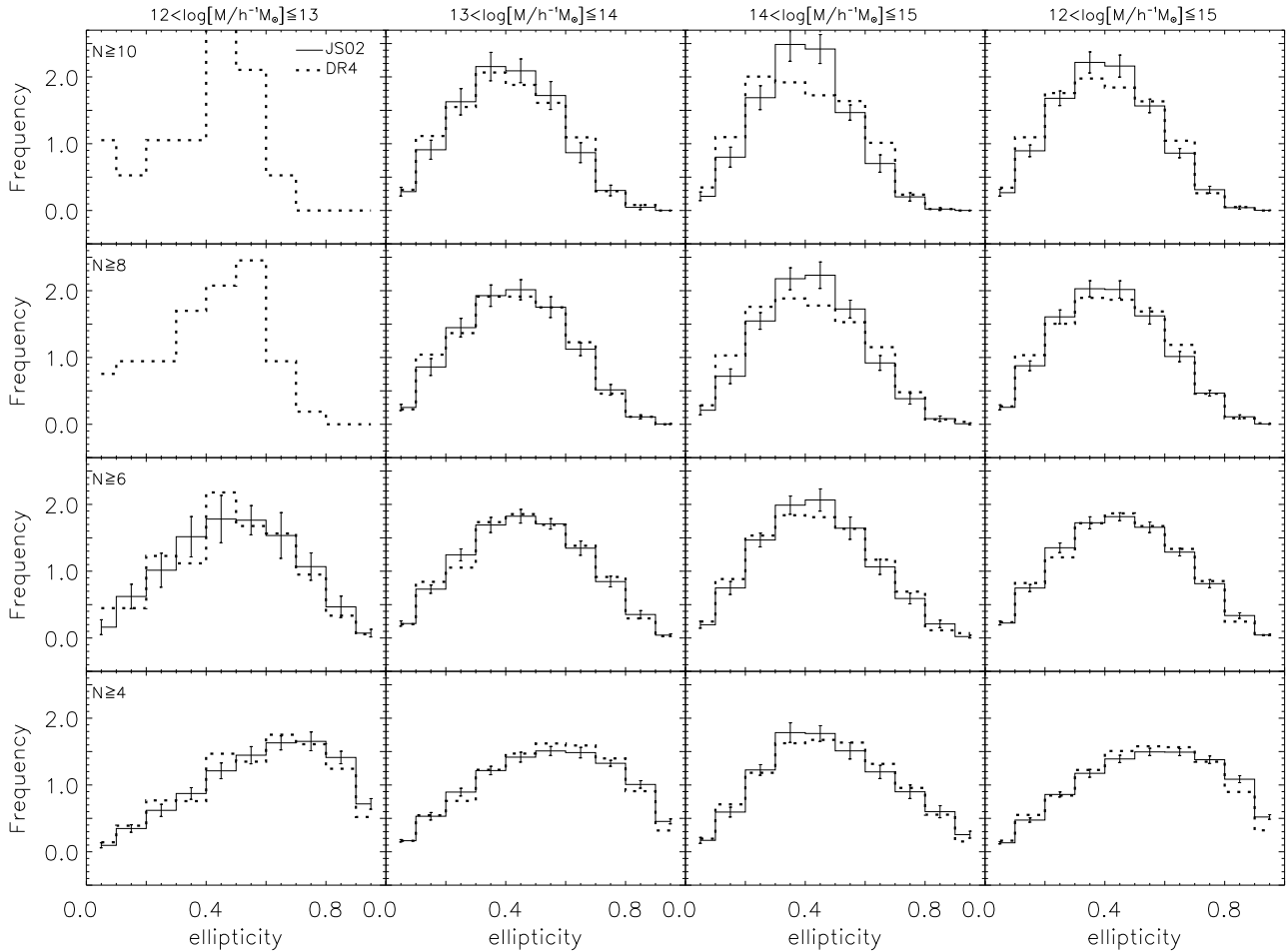}} \caption{Similar to
    Fig.~\ref{fig:NFW},  but  for groups  in  different  mass bins  as
    indicated on  top of  the panels.  In  this plot, the  Monte Carlo
    simulations are  performed according to the JS02  profile with the
    best-fit   three-dimensional  axis   ratios.    The  line-of-sight
    (projection) direction is random.}\label{Fre_tri_bin}
\end{figure*}

\begin{figure*}
%\centerline{\psfig{figure=fig4.ps,width=\hssize}}
  \resizebox{\hsize}{!}{\includegraphics{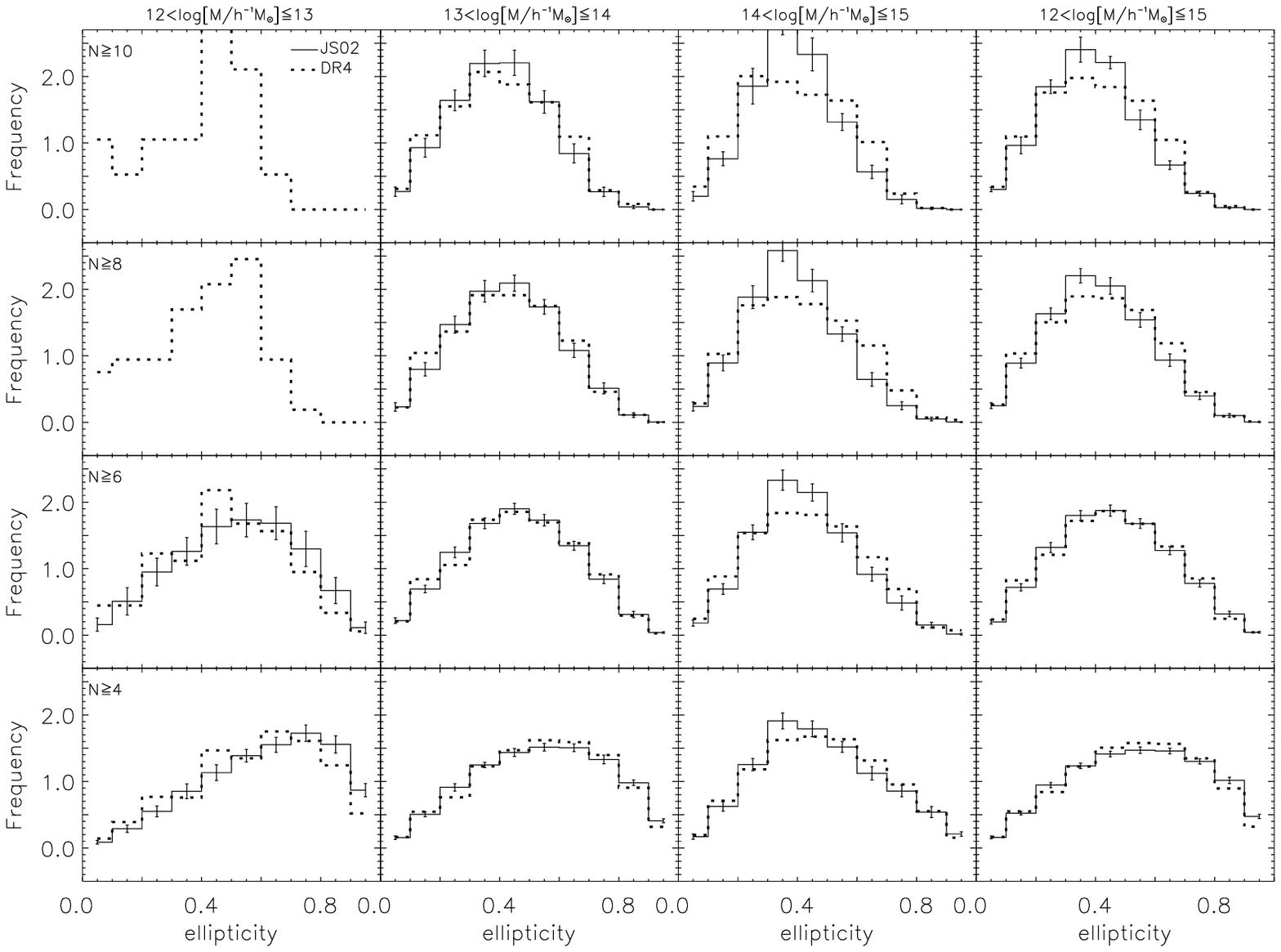}} \caption{Similar to
    Fig.~\ref{Fre_tri_bin}, but for  the best-fit two-dimensional axis
    ratios.}
\label{Fre_2d_bin}
\end{figure*}

In Fig.~\ref{Fre_tri_bin} we  compare the ellipticity distribution
for groups in  different mass bins  (dotted histograms) with those
of the best-fit  Monte-Carlo simulation  (solid histograms  with
errorbars). The corresponding best-fit, three-dimensional (3D) axis
ratios for the dark matter haloes are listed in the second column of
Table~\ref{table:ratios}.  As one can see,  the 3D axis ratios
recovered are different  for groups with different  richness. The
results indicate that less massive haloes are more  spherical, while
more massive haloes  tend to be more prolate.  These results are in
good agreement with Paz et al.  (2006; hereafter P06),  who studied
the shapes of  dark matter  haloes, both projected and in 3D, using
simulations and groups constructed from the 2-degree Field  Galaxy
Redshift Survey (2dFGRS; Colless  et al.  2001) and the SDSS DR3.

In addition  to the three-dimensional axis  ratios, we can also  look into the
projected   two-dimensional   (2D)   axis   ratios.   Assuming   a   projected
major-to-minor  axis  ratio,  we  re-sample  the sky  positions  of  satellite
galaxies in  each group in  a Monte Carlo  way, and measure  the corresponding
ellipticity distribution  for groups of  different masses and  richness.  Here
again, by changing the 2D axis ratio,  we can obtain the one that best matches
the  observed ellipticity  distribution of  groups. The  results are  shown in
Fig.~\ref{Fre_2d_bin},  and the  corresponding  best-fit, 2D  axis ratios  are
listed in  the fourth  column of Table  ~\ref{table:ratios}.  Once  again less
massive groups are found be more spherical.  Note that groups in the same mass
bin,  but  with different  richness,  have  only  slightly different  2D  axis
ratios.

 The $\chi^2$ values presented in Table
  ~\ref{table:ratios} for the best fit models are quite large, especially for
  groups with halo masses $10^{12}-10^{15}\msunh$ in the 3D cases,
  and in a few 2D cases. Possible reasons for such large $\chi^2$ are
  (i) the assumption of a constant axis ratio is not accurate to describe
  the shape of the groups, and (ii) the shapes of observed groups do not follow
  a Gaussian distribution. Unfortunately, without better knowledge
  about these issues, it is difficult to come up with a better model.

Observationally,  only gravitational lensing  data can  directly
probe the 2D, projected mass distribution of dark matter haloes.
However, at the present  time, such observations  are limited in
both  quality and quantity (see  Hoekstra et  al.  2004; Mandelbaum
et al.   2006). For clusters of galaxies, additional data is
available from X-ray data and from  studies  of  the
Sunyaev-Zeldovich-effect.   These  data  yield constraints   on  the
shapes   of  cluster   haloes  from   their  gas distribution.  Mohr
et  al. (1995) used a sample  of 57 X-ray clusters observed by the
Einstein telescope  and obtained a mean 2D axis ratio, $\langle
\eta\rangle$=0.80, and  a  dispersion $\sigma_{\eta}=0.12$. With the
JS02 model, Wang  \& Fan (2004) studied the Sunyaev-Zeldovich effect
and X-ray surface brightness profiles for clusters of galaxies.
Based   on  a   sample   of  clusters   with   masses  above
$M_{\rm
  lim}=10^{14}h^{-1}M_{\odot}$,  they  found  that  the  average  axis
ratios is $\langle \eta \rangle\sim0.84$. Sereno et al.  (2006) used
a sample of  25 X-ray selected  clusters observed by  \emph{Chandra}
and \emph{XMM-Newton}, and  obtained a mean projected  axis ratio
$\langle \eta\rangle =0.80\pm0.02$.  Recently,  Flores et al. (2007)
predicted that the mean axis ratio  is $\langle \eta\rangle =0.82$
and a scatter $\sigma_{\eta}=0.09$ using  a simple analytical model.
Note, though, that  all  these measurements  are  for the axis
ratios  of the  hot intra-cluster  gas.   Lee  \&  Suto (2003)
presented  an  analytical expression that links the ellipticity of
the gas to that  of the dark matter halo,  assuming that the
intra-cluster gas is in hydrostatic equilibrium.  Although this
relation is obtained in 3D, we  assume that it is also valid for the
2D distributions.  By converting the axis ratio of the gas
distribution into the dark matter distribution using  the results in
Fig.~3 of  Lee \& Suto  (2003), we infer 2D axis  ratios for dark
matter haloes of  $0.6\pm 0.1$, in good agreement with  our
measurements ($0.69$  for groups with  halo masses
$10^{14}-10^{15}\msunh$).

\subsection{Dependence of halo shape on the color of central galaxies}

\begin{figure}
%\centerline{\psfig{figure=fig5.ps,width=\hssize}}
\resizebox{\hsize}{!}{\includegraphics{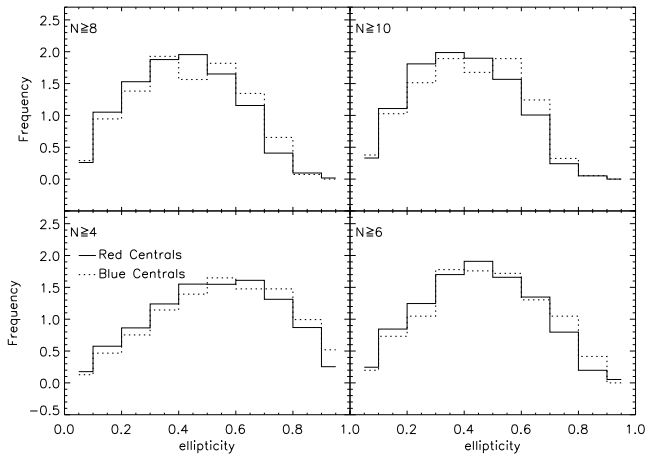}}
\caption{Probability distribution of the ellipticity for groups with
  different  central  galaxies: red  (solid  line)  v.s. blue  (dotted
  line).} \label{Fre_colour}
\end{figure}

The color-magnitude relation  of galaxies is found to  have a
bi-modal distribution, consisting of a red  `sequence' and a blue
`cloud' (e.g. Baldry  et al.2004;  Li et  al.  2006).   In this
subsection  we test whether haloes that  host red or blue central
galaxies have different shapes   (the   motivation   behind   this
will   become   clear   in \S\ref{sec_alignment}).  Following Li  et
al.  (2006) and Yang  et al. (2008) we  separate galaxies into  red
and blue population  using the following dividing curve,
\begin{equation}
^{0.1}(g-r)_0=1.022-0.0651*M_{r,23}-0.00311*M_{r,23}^2\,,
\end{equation}
where  $M_{r,23}=\rmag+23$,  and  $\rmag$  is  the  absolute
magnitude  $K+E$ corrected to $z=0.1$  using the method described in
Blanton \etal (2003).  We define galaxies  with $^{0.1}(g-r)\geq
^{0.1}(g-r)_0$ as red  galaxies and the rest as  blue galaxies. Here
$^{0.1}(g-r)$ is  the color in the  SDSS $g$ and $r$ bands $K+E$
corrected to  $z=0.1$.  In Fig.~\ref{Fre_colour},  we compare the
ellipticity  distribution for groups  with red and blue  central
galaxies. The results indicate that there is no significant
difference between the shape of groups  with blue  or red central
galaxies.

\section{Alignment between central galaxies and dark matter haloes}
\label{sec_alignment}

\subsection{Quantifying the alignment}

In order to quantify the  distribution of satellite galaxies in
groups relative  to the  orientations  of their  central  galaxies
we  follow Brainerd  (2005) and compute  the distribution  function,
$P(\theta)$, where  $\theta$ is the  angle between  the major  axis
of  the central group galaxy and the direction  of a satellite
relative to the central galaxy.   The   angle   $\theta$   is
constrained   in   the   range $0^{\circ}\leq\theta\leq90^{\circ}$,
where      $\theta=0^{\circ} (90^{\circ})$ implies that the
satellite lies along the major (minor) axis of the  central galaxy.
The orientation of  the central galaxy is based on the  isophotal
position angle in the  \emph{r} band, as given in the SDSS-DR4
(Adelman-McCarthy et al. 2006). We have checked the distribution of
these position angles and found it to be isotropic.

For a given set of central  and satellite galaxies, we first count the
total number of central-satellite  pairs, $N(\theta)$, for a number of
bins in $\theta$.   Next, we construct 100 random  samples in which we
randomize  the  orientations  of  all central  galaxies,  and  compute
$\langle N_R(\theta)\rangle$, the  average number of central-satellite
pairs as function of $\theta$.  Note that this ensures that the random
samples have exactly the same selection effects as the real sample, so
that any significant  difference between $N(\theta)$ and $N_R(\theta)$
reflects a  genuine alignment between the orientations  of the central
galaxies  and  the  distributions  of  their  corresponding  satellite
galaxies.

To quantify the  strength of any possible alignment  we follow Y06 and
define the distribution of normalized pair counts:
\begin{equation}\label{eq:fpairs}
f_{\rm pairs}(\theta)=\frac{N(\theta)}{\langle N_R(\theta)\rangle}.
\end{equation}
Note that in the  absence of any alignment, $f_{\rm pairs}(\theta)=1$,
while $f_{\rm pairs}(\theta)>1$ at  small $\theta$ implies a satellite
distribution with a preferred alignment  along the major axis of their
central   galaxy.  We   use  $\sigma_{\rm   R}(\theta)/\langle  N_{\rm
  R}(\theta)\rangle$, where $\sigma_{\rm R}$ is the standard deviation
of $N_{\rm R}(\theta)$ obtained from the 100 random samples, to
assess the  significance of  the  deviation of  $f_{\rm
pairs}(\theta)$  from unity. In addition to this  normalized pair
count, we also compute the average angle $\langle\theta\rangle$. In
the absence of any alignment $\langle\theta\rangle =45^{\circ}$,
however, $\langle\theta\rangle=45^{\circ}$   does   not   mean   an
isotropic distribution.  Major  and minor  axis alignments are
characterized by $\langle\theta\rangle<45^{\circ}$ and
$\langle\theta\rangle>45^{\circ}$, respectively.   The significance
of any  alignment can  be expressed  in terms  of $\sigma_{\theta}$,
the variance in  $\langle\theta\rangle_R$, which is obtained  from
the 100 random samples.

\begin{figure}
%\centerline{\psfig{figure=fig6.ps,width=\hssize}}
\resizebox{\hsize}{!}{\includegraphics{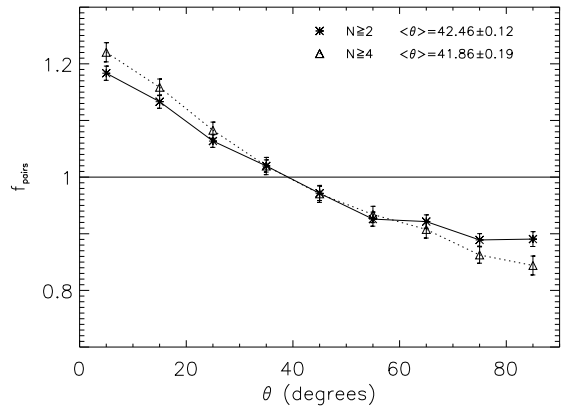}}
\caption{The normalized probability distribution, $f_{\rm
    pairs}(\theta)$, of  the angle $\theta$ between the  major axis of
  the central galaxy and the direction connecting the satellite galaxy
  and the  central galaxy.   Results are measured  for groups  with at
  least  two members  (solid line)  and four  members (dotted  line).}
\label{fp_all}
\end{figure}

Fig.~\ref{fp_all}    shows    $f_{\rm    pairs}(\theta)$    for all
central-satellite  pairs (solid  line) in  our SDSS  group
catalogue. Clearly $f_{\rm pairs}(\theta)>1$  for small $\theta$,
indicating that satellite galaxies are distributed preferentially
along the major axis of  their central galaxy.   This is  also
evident  from the  fact that
$\langle\theta\rangle=42.46^{\circ}\pm0.12^{\circ}$,   which
deviates from  the  case  of  no  alignment  (i.e.,  $\langle\theta
\rangle  = 45^{\circ}$) by almost $21\sigma$!   For comparison,
using the a group catalogue constructed from the SDSS DR2 data by
Weinmann \etal (2006), similar   to   that  used   here,   Y06 found
$\langle\theta\rangle =42.2^{\circ}  \pm  0.2^{\circ}$,  in
excellent  agreement  with  the results  presented here.   Note,
though, that  the statistical  error presented  here is  much
smaller,  due to  the larger  group catalogue used.  The  existence
of alignment  owns in part to  the non-spherical distribution of the
satellite galaxies  in dark matter  haloes (e.g., Zentner et al.
2005; Kang et al.  2005, 2007; Libeskind et al.  2005; AB06), which
has been used in the previous  section to probe the  overall shapes
of  the dark matter mass   distribution.    For    comparison,   the
dotted   lines   in Fig.~\ref{fp_all}  show  $f_{\rm pairs}(\theta)$
for  groups with  at least  four  members.   The  resulting
alignment  signal  is  slightly stronger   than   for   the    full
sample   (which   includes   all central-satellite pairs in groups
with  at least two members). This is consistent with the fact that
(i) groups with more satellites are more massive  and (ii)  more
massive  groups are  less spherical  (see also
\S\ref{sec:align_mass} below).

\subsubsection{Dependence on galaxy color}

\begin{figure}
%\centerline{\psfig{figure=fig7.ps,width=\hssize}}
\resizebox{\hsize}{!}{\includegraphics{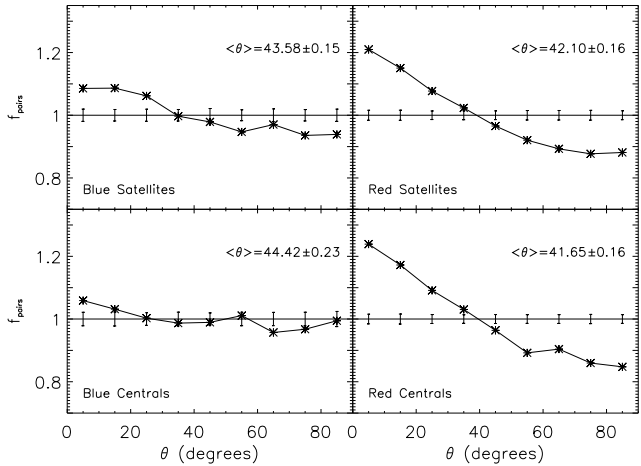}}
\caption{Same as Fig.~\ref{fp_all}, but for different subsamples  of
  central-satellite  pairs, separated  according to  the $^{0.1}(g-r)$
  colors  either for  satellite galaxies  (upper two  panels)  and for
  cental galaxies (lower two panels).}
\label{fp_cs1}
\end{figure}

In order to  study how the alignment depends  on various properties of
the  central and  satellite galaxies,  we  follow Y06  and divide  our
sample  of  central-satellite pairs  into  different subsamples.   The
upper panels of Fig.~\ref{fp_cs1}  show the alignment signals obtained
for blue and  red satellites, while the lower  panels show the results
for  blue  and red  centrals.   As  one can  see,  there  is a  strong
dependence  on the  colors of  both the  centrals and  satellites.  In
particular,  groups  with  red  centrals  and red  satellites  show  a
stronger alignment than those  with blue centrals and blue satellites,
in  good agreement  with  previous studies  (Y06;  Azzaro \etal  2007;
Agustsson \& Brainerd 2007).  As  pointed out by K07, groups with blue
centrals tend  to have slightly  more interlopers, which may  cause an
arteficial reduction of their measured alignment.  However, even after
this is  corrected for,  groups with blue  centrals still show  a weak
alignment.

\begin{figure}
%\centerline{\psfig{figure=fig8.ps,width=\hssize}}
\resizebox{\hsize}{!}{\includegraphics{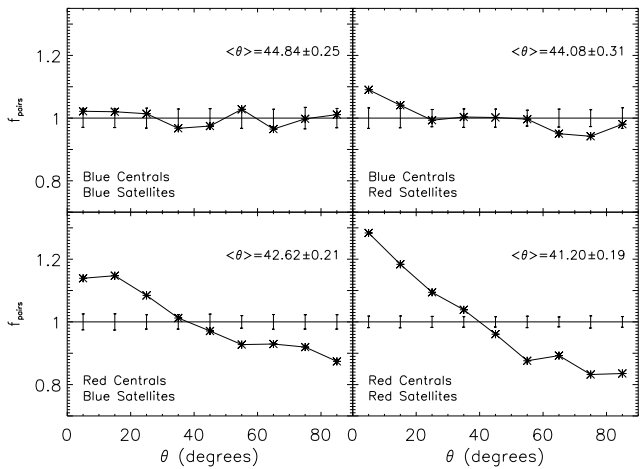}}
\caption{Same  as Fig.~\ref{fp_cs1}, except that here  we split the
  sample  according  to  different  combinations  of  the  colors  for
  \emph{both} central and the satellite galaxies, as indicated.}
\label{fp_cs2}
\end{figure}

In Fig.~\ref{fp_cs2} we show  the alignment , $f_{\rm pairs}(\theta)$,
for the  four color combinations  of centrals and satellites.   As one
can see, pairs  between blue centrals and blue  satellites do not show
any  alignment, while pairs  between red  centrals and  red satellites
show  the strongest alignment.   Pairs between  red centrals  and blue
satellites  and pairs between  blue centrals  and red  satellites show
alignment  with  intermediate strength.   All  these  findings are  in
excellent agreement with, and more significant than, those obtained in
Y06.

\subsubsection{Dependence on halo mass}\label{sec:align_mass}

\begin{figure}
%\centerline{\psfig{figure=fig9.ps,width=\hssize}}
  \resizebox{\hsize}{!}{\includegraphics{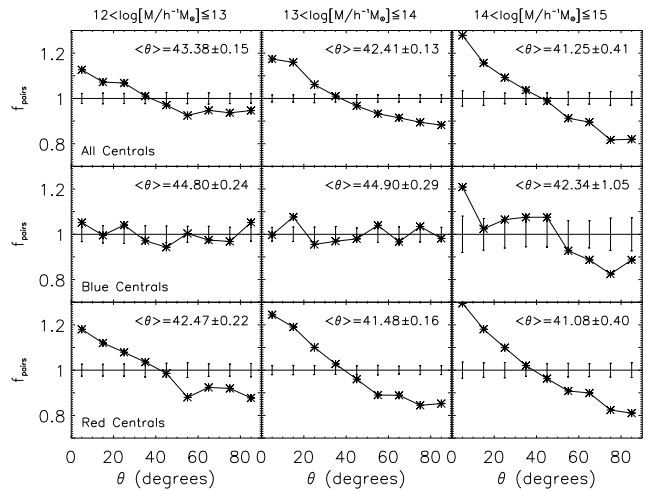}}
\caption{Similar to Fig.~\ref{fp_all}, but for central-satellite pairs
  in haloes  of different masses  as indicated on  top of the  panels.
  Results are shown separately for  all, red and blue central galaxies
  in the upper, middle and lower panels.}
\label{fp_mass}
\end{figure}

Fig.~\ref{fp_mass} shows the alignment measure for groups of
different halo masses.  From the upper panels  one can see that the
alignment is stronger for more massive groups.  We also examine the
mass dependence of the alignment separately for groups with blue and
red centrals, the results  of  which  are  shown  in  the middle and
lower  panels  of Fig.~\ref{fp_mass},  respectively.  The alignment
is  quite different for  blue and  red  centrals. For  all halo
masses  probed here,  red centrals  reveal  a  much  stronger
alignment  with  their  satellite galaxies  than blue  centrals. In
fact, except  for the  most massive haloes  with $14\le  \log
M/\msunh  \le  15$, we  find no  significant alignment signal  of
satellites with blue centrals.   These results are again in good
agreement with  the findings  of Y06  based  on galaxy groups in the
SDSS DR2 (see also Agustsson \& Brainerd 2007).

\subsection{The alignment of central galaxies with their host haloes}

\begin{figure}
%\centerline{\psfig{figure=fig10.ps,width=\hssize}}
\resizebox{\hsize}{!}{\includegraphics{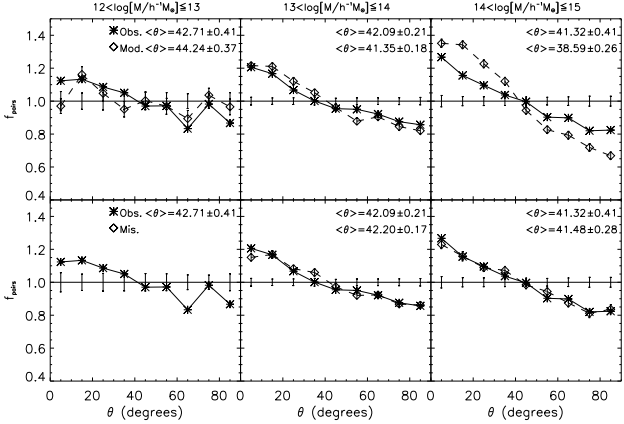}}
\caption{The alignment signal $f_{\rm pairs} (\theta)$ obtained by
  consider the  misalignment-angle between the major axis  of the dark
  matter halo and  the major axis of the  central group galaxy.  Upper
  panels:  the alignment  signals by  assuming that  the  direction of
  central galaxy is same as  the direction of the projected major axis
  of  the dark  matter halo  (dashed line),  comparing to the observed
  alignment  signal  (solid  line).   Lower  panels:  The  comparisons
  between  the observed  alignment signal  (solid line)  and  the best
  fitting alignment  signal by considering a  misalignment between the
  projected major axis  of the dark matter halo and  the major axis of
  the central galaxy (dashed line).}
\label{fp_mis}
\end{figure}

As shown above, central galaxies  are aligned with the distribution of
their satellite galaxies, and  different systems show different levels
of alignment.  To  produce such an alignment signal  requires that (i)
dark matter  haloes are elongated,  (ii) satellite galaxies  trace the
dark matter, at least to some extent, and (iii) the orientation of the
central galaxy is  correlated with the orientation of  its dark matter
halo.  Now  that we  have obtained constraints  on both the  shapes of
dark  matter haloes,  we can  use  the observed  alignment signals  to
constrain the correlation between the orientations of central galaxies
and their dark matter haloes.

Previously,  such  correlations  have  been constrained  using  galaxy
catalogues  constructed from semi-analytical  models (SAM)  for galaxy
formation (see AB06 and K07).  In K07, the satellite distributions are
modelled using the  locations of subhaloes (which are  thought to host
the  satellite  galaxies)  inside  larger  dark  matter  haloes  in  a
numerical simulation.  In general,  subhaloes are found to be accurate
tracers of the  shapes of the dark matter  distributions of their host
haloes.  However, since the SAM do not predict the orientations of the
central galaxies,  a number  of simple assumptions  have been  made so
far, where the minor axis of the central is perfectly aligned with (i)
the major axis of the inertia  tensor of the host halo, (ii) the minor
axis of  the inertia tensor of  the host halo,  (iii) the intermediate
axis  of the  inertia tensor  of the  host halo,  or (iv)  the angular
momentum  vector of  the host  halo.  Both  AB06 (using  the alignment
signal measured  from isolated host-satellite systems)  and K07 (using
central-satellite  pairs  in  galaxy  groups)  found  that  model  (i)
predicts an alignment signal that is much too strong compared with the
data, model (ii)  predicts a strong Holmberg effect,  contrary to what
is seen,  and model  (iii) predicts almost  no alignment.   Only model
(iv)  results  in  alignment   signals  that  are  in  agreement  with
observations.   Although this does  not give  definite proof  that the
minor axis of the central galaxy is perfectly aligned with the angular
momentum vector of the halo,  it does provide a possible understanding
for  the origin  of a  correlation between  the orientation  a central
galaxy and that of its host halo.

However, it is important to keep  in mind that the SAM not
necessarily predict  the  correct spatial  distributions  of
satellite galaxies. Therefore, in  this paper we  focus on what  can
be inferred  from the data alone.  In particular, we try to {\it
infer}, based purely on the data presented  here, to what extent the
{\it projected} orientations of  central galaxies  are  aligned with
those  of  their dark  matter haloes.  Similar to what we have done
to extract the three-dimensional and two-dimensional axis ratios in
Section~\ref{sec_shape}, we compare the data to Monte Carlo
simulations.  We start from the groups with at least four members,
and assume that the projected  orientations of the central galaxies
are perfectly aligned with the orientation of the projected
satellite distribution. Next, using a Monte Carlo method, we
distribute the  satellite galaxies according to the projected
two-dimensional   axis  ratios listed in the fourth column of
Table~\ref{table:ratios} for the group sample with $N\geq4$.
Finally, we  measure  the  various alignment signals for this Monte
Carlo simulation using the same method as described above. The upper
panels of Fig.~\ref{fp_mis} show the alignment signals thus obtained
(open symbols) for groups of different masses, as indicated  at  the
top  of  each  panel. For comparison, we also plot the data as
asterisks. Clearly, this model predicts  an alignment  signal  that
is  stronger  than observed, especially for the  more  massive
groups.

We can suppress the strength of  the alignment signals in the model by
taking scatter (i.e., `random' deviations from perfect alignment) into
account.  To that extent we assume that the misalignment angle between
the projected orientation of the  central galaxy and the major axis of
its host halo can be described by a Gaussian distribution:
\begin{equation}
p(\theta_{\rm mis})=\frac{1}{\sqrt{2\pi}\sigma}\exp
\left( {-\frac{\theta_{\rm mis}^2}{2\sigma^2}}\right) \,,
\end{equation}
with $\sigma$  the standard deviation of the  distribution.  Note that
this  distribution for  the misalignment  angle $\theta_{\rm  mis}$ is
symmetric and centered around zero.

Using  such a  Gaussian distribution,  we  fit the  various
alignment  signals obtained from our SDSS group  catalogue, treating
$\sigma$ as a free parameter in each separate case. Here again  we
use the minimum $\chi^2$ fit, similar to Eq. \ref{eq:chi2} but for
the angular distribution of central-satellite pairs (Eq.
\ref{eq:fpairs}). The parameter  $\sigma$ is  constrained in  the
range $0^{\circ}-90^{\circ}$, and on  the regular 900 grids. For
each $\sigma$, the misalignment angles $\theta_{\rm mis}$ are
generated according to the Gaussian distributions. We measure the
alignment signals in Monte Carlo mock samples in the same way as the
observations, and obtain $\chi^2$ values on 900 grids. The best-fit
parameter $\sigma$  can  be obtained  from  the grid  which has  the
minimum  $\chi^2$ value. As an illustration,  in  the  lower panels
of Fig.~\ref{fp_mis}  we compare  the alignment  signals obtained
from the data (asterisks) with those obtained from the Monte Carlo
simulations including the best-fit Gaussian distribution of
misalignment angles (open symbols). Clearly, the observed results
can be well reproduced by such a model. Table~3 lists the best-fit
values  of $\sigma$ thus  obtained, for a  variety of cases  that
are shown  in  Figs.~\ref{fp_all}  -   \ref{fp_mass}.  To  better
describe  these statistical values,  $68.3\%$ confidence levels are
also  given following each best fit $\sigma$. Here the confidence
levels are obtained from the grids with $\chi^2=\chi^2_{min}  +
1.0$.  Clearly,  the   amount  of  scatter   in  the misalignment
angle depends strongly on the  group mass and on the color of the
central  galaxy.  For  the  entire sample  as  a whole  we obtain
$\sigma\sim 23^\circ$,  while   groups  with  red  centrals,  on
average,  have  smaller misalignment  angles  than those  with  blue
centrals.  Owing to  the  nearly spherical shapes of dark matter
haloes with $12.0< \log [M/\msunh] \le 13.0$, the misalignment angle
could not be meaningfully constrained in these cases.

In a recent study, K07 found that  if the minor axis of the central
galaxy and the angular momentum  vector of the dark matter halo  is
in perfect alignment, most  of   the  observational
satellite-central  alignment   signals  can  be automatically
reproduced. On the other  hand, any processes that can introduce an
average  misalignment $\sim 40^{\circ}$  between the minor axes  of
central galaxy  and host  dark matter  halo can  also explain  the
observed alignment signals.  Using high  resolution  numerical
simulations,  Bailin \&  Steinmetz (2005) found that  the angular
momentum vectors  tend to align  with the minor axes of the dark
matter halo  with a mean misalignment of $\sim 25^{\circ}$ at small
radii  and of $\sim  40^{\circ}$ at the  halo virial radius. Here to
be able to  compare with  our findings, as  an experiment, we input
misalignment angles (with an average $\sim 40^{\circ}$ and some
scatters) between the minor axes of the  dark matter haloes and the
central  galaxies. By randomly project the 3D shapes  of the dark
matter haloes and central  galaxies, we measure the distribution of
the angles between the  {\it projected major} axes of the dark
matter halos and central galaxies,  if modelled with an Gaussian
distribution, which roughly corresponds  to $\sigma \sim
22^{\circ}$. Thus  our finding are in remarkably good agreement with
K07 and Bailin \& Steinmetz (2005).

\begin{table}
 \caption{Best fitting parameters for the deviation angle between the
   orientation
of the central galaxy and the dark matter halo.}\label{msi_ali}
\begin{center}
\begin{tabular}{lcccc}\hline
 Subsample &$\sigma$ (degree)\\
 \hline\hline

  all$^a$              & $23.3^{+0.1}_{-0.3}$\\
  blue cen.            & $38.1^{+0.1}_{-0.2}$\\
  red cen.$^b$         & $16.6^{+0.1}_{-0.1}$\\
  blue sat.            & $29.3^{+0.8}_{-0.1}$\\
  red sat.$^c$         & $19.9^{+0.7}_{-0.1}$\\
  red cen./blue  sat.  & $38.5^{+0.1}_{-0.6}$\\
  red cen./red   sat.  & $16.1^{+0.3}_{-0.1}$\\
  blue cen./blue sat.  & $80.0^{+7.8}_{-21.4}$\\
  blue cen./red  sat.  & $37.9^{+0.2}_{-0.1}$\\
  red  cen. (M23$^d$)   & --\\
  red  cen. (M34)       & $7.6^{+0.3}_{-0.1}$\\
  red  cen. (M45)       & $25.7^{+0.3}_{-0.1}$\\
  blue cen. (M23)       & --\\
  blue cen. (M34)       & $64.8^{+0.1}_{-0.5}$\\
  blue cen. (M45)       & $27.2^{+0.1}_{-0.1}$\\
  all (M23)        & --\\
  all (M34)        & $23.3^{+0.2}_{-0.3}$\\
  all (M45)        & $30.2^{+0.1}_{-0.2}$\\
 \hline
 \end{tabular}
\end{center}
 {\footnotesize
 \noindent
 $^{a}$ all means all central galaxies\\
 $^{b}$ blue cen. and red cen. denote blue centrals and red centrals,
        respectively. \\
 $^{c}$ blue sat. and red sat. denote blue satellites and red satellites,
        respectively.\\
 $^{d}$ M23 means that the halo masses are in the range
        $12<{\rm log}[M/\msunh]\le13$.
        Note that in M23, the alignment signals for the perfect
        alignment are already as weak as the observed ones,
        and so no misalignment angle is introduced. \\
}
\end{table}

\section{Summary and discussion}

Using the large galaxy group catalogues constructed from the SDSS
Data Release  4 (DR4) by  Yang et  al.  (2007),  we have
investigated the shapes of their  host dark matter haloes, and  the
correlation between the  orientations of  the central  galaxies  and
those  of their  host haloes.    In  particular,   we  obtained
the   two-dimensional  and three-dimensional  axis  ratios  of
galaxy groups  by  comparing  the observed,  projected ellipticity
distributions of  satellite galaxies with  those  of  Monte   Carlo
simulations,  and  we  determined  the probability  distributions
for the  angles between  the major  axis of central  galaxies and
the lines  connecting the  centrals  with their satellites. The main
results of this paper are summarized as follows.

\begin{enumerate}
\item Under the assumption  that the spatial distribution of satellite
  galaxies traces the  shapes of the underlying dark  matter haloes we
  find that the projected ellipticity distributions are slightly better
  fit with the triaxial models of JS02 than with simple, spherical NFW
  models.
\item The shapes of dark matter haloes depend strongly on their mass, with
  more massive haloes being more elongated.  Haloes with masses in the range
  $12<{\rm log}[M/\msunh]\leq13$ are nearly spherical, while more massive
  haloes with $14<{\rm log}[M/\msunh]\leq15$ are more prolate.
\item There is  no significant difference between the  shapes of haloes
  with red central galaxies and those with blue central galaxies.
\item Satellites  are preferentially distributed along  the major axes
  of their  central galaxies. The  strength of this  alignment depends
  strongly  on  halo  mass and  on  the  colors  of both  central  and
  satellite galaxies. The alignment  is strongest between red centrals
  and red satellites, while blue  centrals show almost no alignment at
  all. More massive groups show a stronger alignment than less massive
  groups.
\item  The  observed alignment  can  be  reproduced  if the  projected
  orientation  of  central  galaxies  is  aligned  with  that  of  the
  projected mass  distribution of its halo. However,  the alignment is
  not perfect.  The  data can be reproduced under  the assumption that
  the misalignment angle follows  a Gaussian distribution around zero,
  and with a standard deviation  of $\sim 23$ degrees. Because of (ii)
  and (iii)  and (iv), groups with  blue centrals have,  on average, a
  larger misalignment angle than those with red centrals.
\end{enumerate}

Our findings regarding the shapes of the galaxy groups are in
qualitative agreement with those of P06, and Plionis et al.  (2004).
However, there are quantitative difference, especially for small
haloes.  We find that small haloes are almost spherical, and the
axis ratios (major-to-minor) in our measurements are much smaller
than their measurements.  Note that we have used Monte Carlo
simulations to infer the shapes of dark matter haloes, which avoids
the impact of selection effects, while in Plionis et al.(2004) the
axis ratios are directly measured from the distribution of member
galaxies. Although they found that poor groups are more elongated
than rich ones, they cautioned that their results may be
significantly affected by discreetness effects (see also Plionis et
al. 2006). Numerical simulations (e.g, Kasun \& Evrard 2005; Allgood
et al.  2006) have shown that the small haloes are more spherical
than massive haloes, in good agreement with our findings here.

The alignment signals presented here are in good agreement with
those obtained by Y06 using a similar, but  smaller group catalogue
constructed from the SDSS DR2, and with other studies (Brainerd
2005; Agustsson \& Brainerd 2006a, 2007; Donoso,  O'Mill  \&  Lambas
2006;  Faltenbacher  \etal  2007a;  Azzaro  \etal 2007). Contrary to
the previous studies by AB06 and K07,  our analysis of the
implications  for  the correlation  between  the  orientation  of
the  central galaxies and that of its halo are based purely on the
data presented here, and therefore  does  not  depend  on  any
galaxy  formation  models  or  numerical simulations. Nevertheless,
tests show that  their models of the  perfect alignment between the
minor axis of the  central galaxy and angular  momentum vector of
dark matter  halo are  in good  agreement with our  direct
measurement  of the projected misalignment angles.

%%%%%%%%%%%%%%%%%
% Ackowledgements
%%%%%%%%%%%%%%%%%

\section*{Acknowledgments}
We sincerely thank the referee Manolis Plionis for the constructive
and detailed comments and suggestions. We also acknowledge helpful
discussions with Yipeng Jing, Changbom Park and Xi Kang. This
research was supported by the National Science Foundation of China
under  grants 10373001, 10533010, 10533030, 10673023, 10773001, the
Distinguished Young Scholar Grant 10525314, by the Chinese Academy
of Sciences under grant KJCX3-SYW-N2, KJCX2-YW-T05, and the {\it One
Hundred Talents} project, by the Shanghai Pujiang Program (No.
07pj14102), by the 973 Program (No. 2007CB815401, 2007CB815402), by
the NSF under grant AST-0607535, IIS-0611948, and by the NASA under
grant AISR-126270.

\end{document}